\documentclass[preprint,aps]{revtex4}
\usepackage{epsfig}
\usepackage{latexsym}

\begin{document}

\title{Self-Organized Dynamical Equilibrium in the Corrosion of 
  Random Solids}

\author{Santanu Sinha,  Vimal Kishore and S. B. Santra}

\affiliation{Department of Physics, Indian Institute of Technology
Guwahati, Guwahati-781039, Assam, India.}


\begin{abstract}
Self-organized criticality is characterized by power law correlations
in the non-equilibrium steady state of externally driven systems. A
dynamical system proposed here self-organizes itself to a critical
state with no characteristic size at ``dynamical equilibrium''.  The
system is a random solid in contact with an aqueous solution and the
dynamics is the chemical reaction of corrosion or dissolution of the
solid in the solution. The initial difference in chemical potential at
the solid-liquid interface provides the driving force. During time
evolution, the system undergoes two transitions, roughening and
anti-percolation. Finally, the system evolves to a dynamical
equilibrium state characterized by constant chemical potential and
average cluster size. The cluster size distribution exhibits power law
at the final equilibrium state.
\end{abstract}

\maketitle

The phenomenon that a class of externally driven system evolves
naturally into a state of no single characteristic size or time is
known as self-organized criticality (SOC)\cite{jensen}. SOC is
observed in several situations like sandpile\cite{bak},
earthquakes\cite{earthq}, forest fire\cite{ff},
coagulation\cite{coag}, river networks\cite{rnetw}, etc.  The models
demonstrating SOC are in out of equilibrium situations. The
non-equilibrium steady state in SOC is characterized by long range
spatio-temporal correlations and power law scaling behaviour.

\begin{figure}
\centerline{\psfig{file=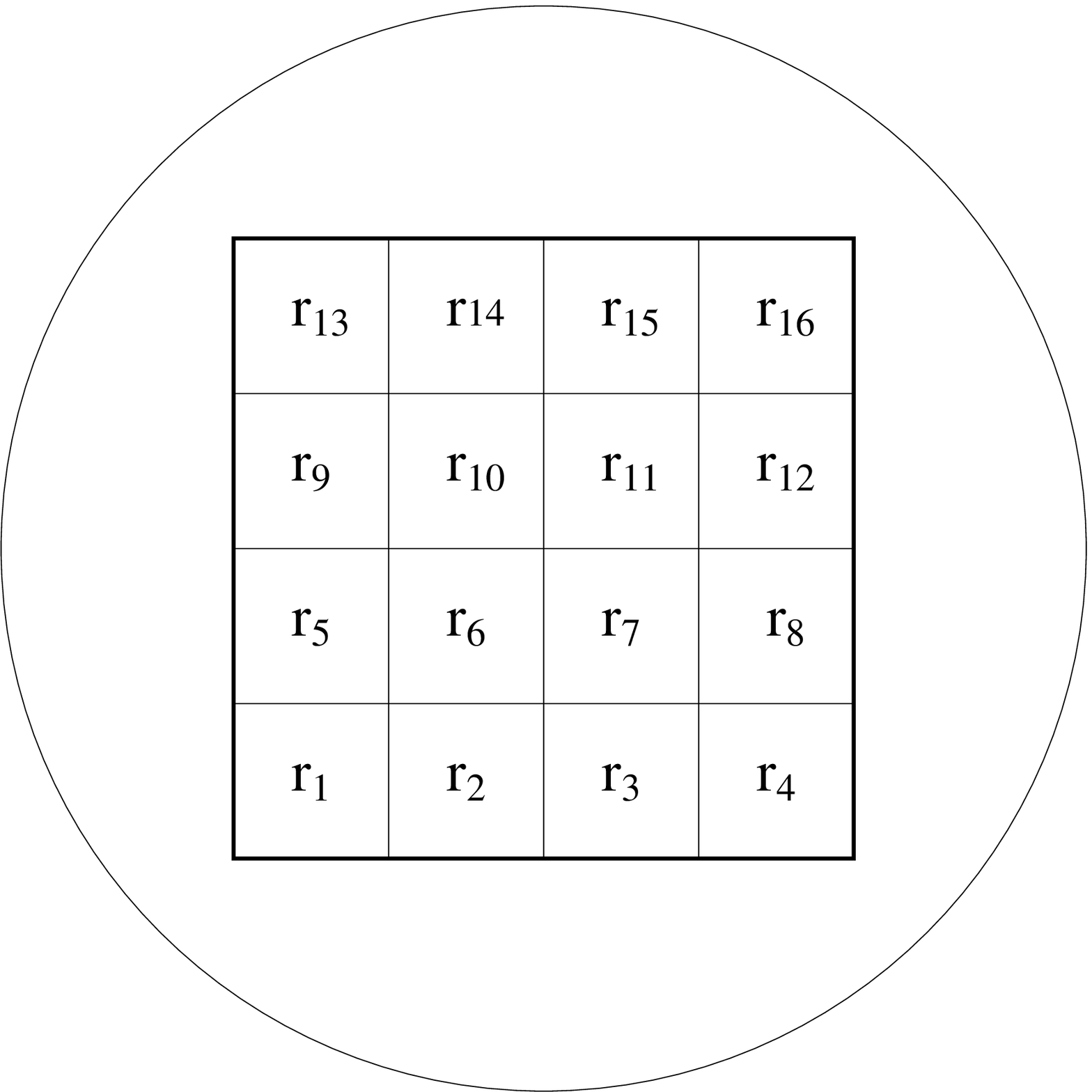,width=0.32\textwidth} \hfill
\psfig{file=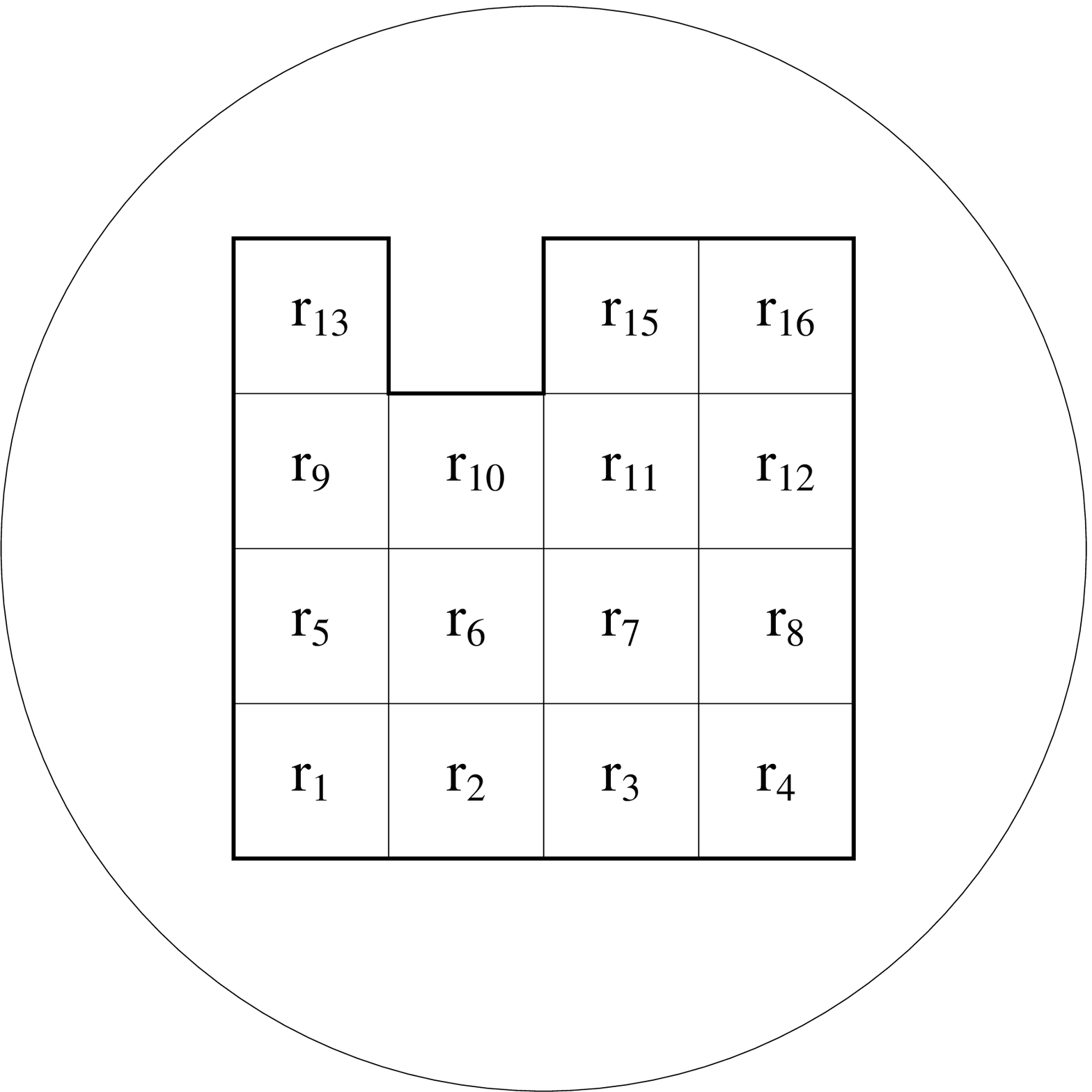,width=0.32\textwidth}\hfill
\psfig{file=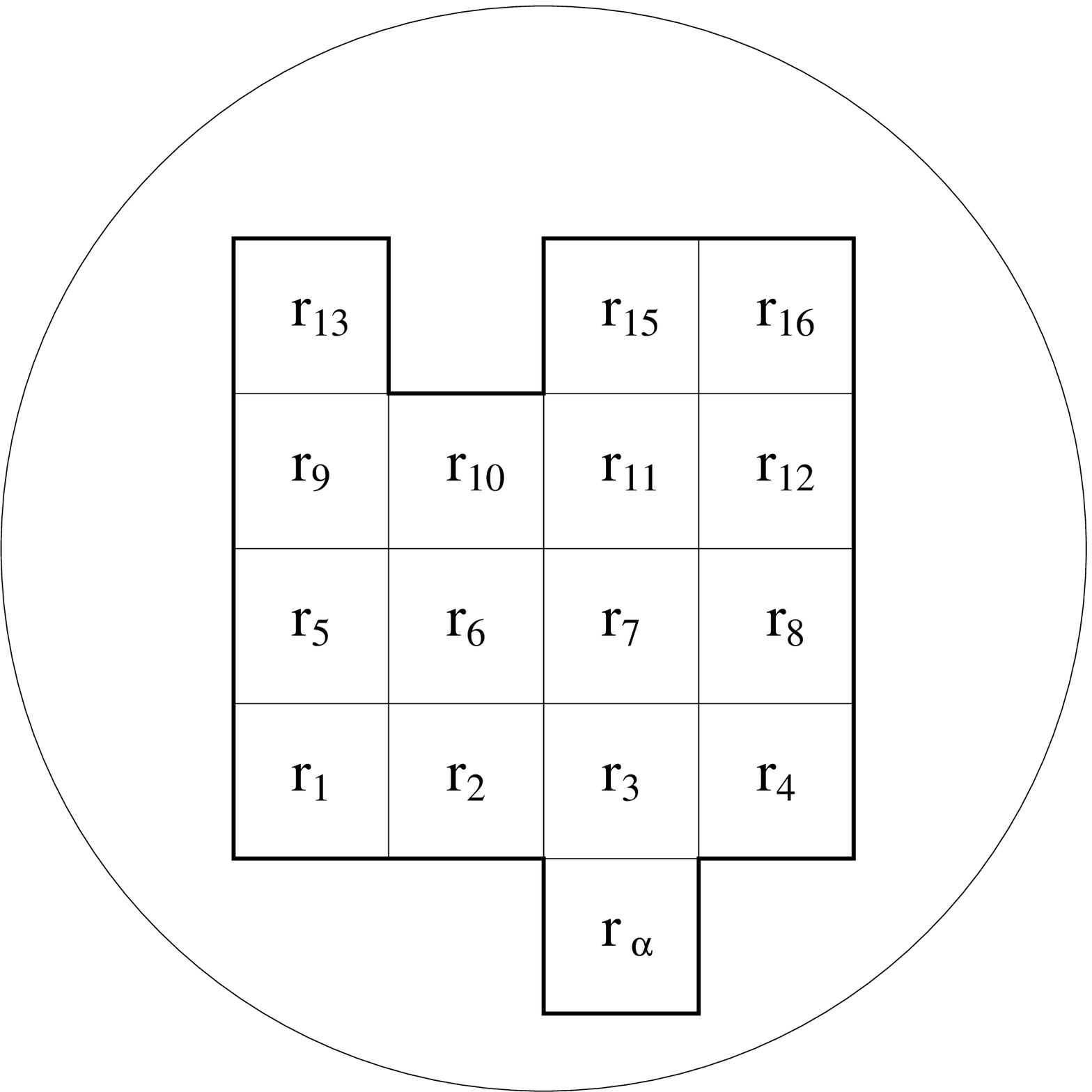,width=0.32\textwidth} }
\centerline{\hfill $(a)$ \hfill \hfill $(b)$ \hfill\hfill $(c)$\hfill}
\caption{\label{demo} A single Monte Carlo (MC) step is represented
  here. $(a)$ The arrangement of random numbers $r_i$ represents a
  random solid, $(b)$ $r_{14}$ is identified as minimum random number
  on the external boundary of the solid and it is dissolved, $(c)$ the
  solid element is redeposited at a randomly chosen site $r_3$ of the
  modified solid surface. A new random number $r_\alpha$ is associated
  with the redeposited site. The process is repeated.}
\end{figure}

In this paper, a system is demonstrated which is evolving naturally
into a dynamical equilibrium state with no characteristic size. The
system considered here is the corrosion or dissolution of a random
solid in an aqueous solution. The initial difference in the chemical
potential of the solid-liquid interface drives the system and no other
external force is applied. A random solid could be a multicomponent
vitreous system in which, due to random local chemical environment,
the binding energies of the constituent molecules of the solid are
expected to be arbitrary. In order to study the time evolution of such
a system in aqueous solution, a numerical model of corrosion or
dissolution is developed in $2$ dimensions ($2D$). In this model, it
is assumed that the random binding energy is uniformly distributed
between $0$ and $1$. The solid shown in figure \ref{demo} ($a$) is a
dense structure of elements with random binding energy $r_i$. The
solid is placed in an aqueous solution, assumed to be infinite. The
white space in figure \ref{demo} represents the solution. Random solid
element $R$ dissolves slowly in the solution $S$, makes a compound
$RS$ and the compound breaks instantaneously into $R$ and $S$. The
chemical reaction of dissolution could be represented as
\begin{equation}
\label{chr}
R + S \rightarrow RS \rightarrow R + S. 
\end{equation}
The solid element $R$ in the solution is now available for
redeposition on the interface. Assuming diffusion of the solid element
$R$ is very fast in the solution, redeposition is made at a randomly
chosen site on the externally accessible perimeter with unit
probability. Generally, dissolution is a slow process and redeposition
is faster. It is mimicked here by considering no further dissolution
during redeposition. The slowest possible dynamics of the system then
involves dissolution and redeposition processes together with
reconstruction of the rough interface at the single particle
level. Different numerical processes involved in a single Monte Carlo
(MC) step are explained with the help of figure \ref{demo}. The
process are: $(i)$ extraction of externally accessible perimeter of
the solid, $(ii)$ dissolution of the site with minimum random number
(minimum binding energy) on the perimeter with unit probability (in
figure \ref{demo} ($b$), it is $r_{14}$), $(iii)$ modification of the
external perimeter, $(iv)$ redeposition of the solid element present
in the solution on a randomly chosen site of the modified external
perimeter ($r_3$ in figure \ref{demo} $(c)$), $(v)$ assignment of a
new random number $r_\alpha$ to the redeposited site. The whole
process is then repeated and time is increased from $t$ to $t+1$. Note
that the total number of particles $(L^2)$ is conserved and the system
evolves at equal solid to liquid and liquid to solid flux rate (one
particle per time step) throughout the simulation.

\begin{figure}
\centerline{\hfill\psfig{file=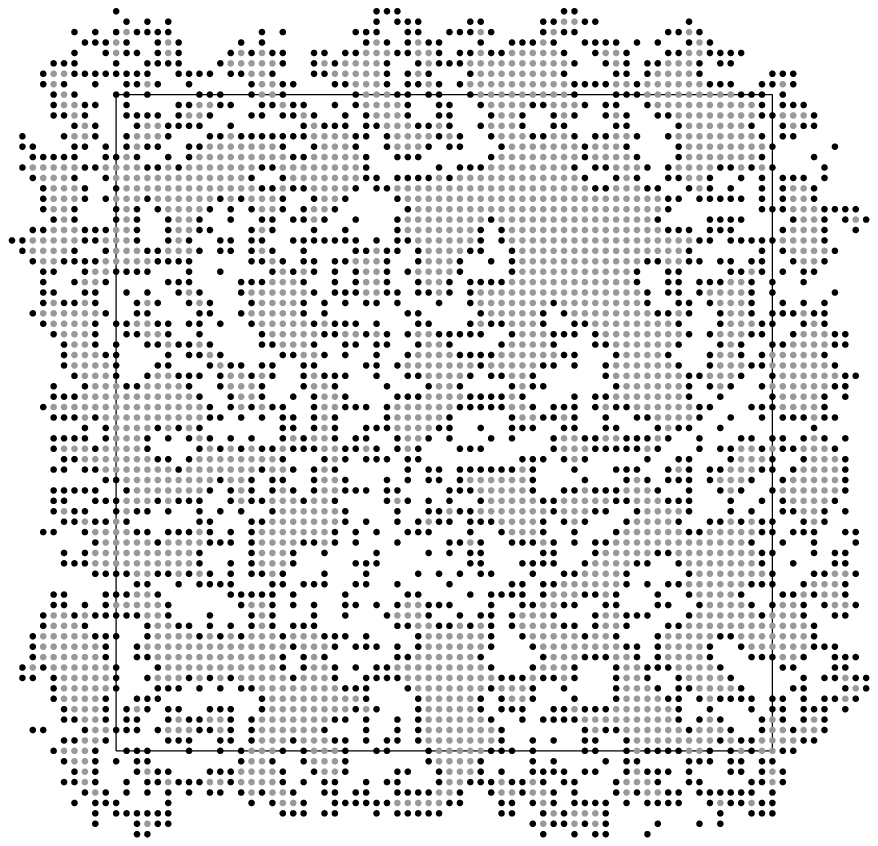,width=0.32\textwidth}
  \hfill\hfill\psfig{file=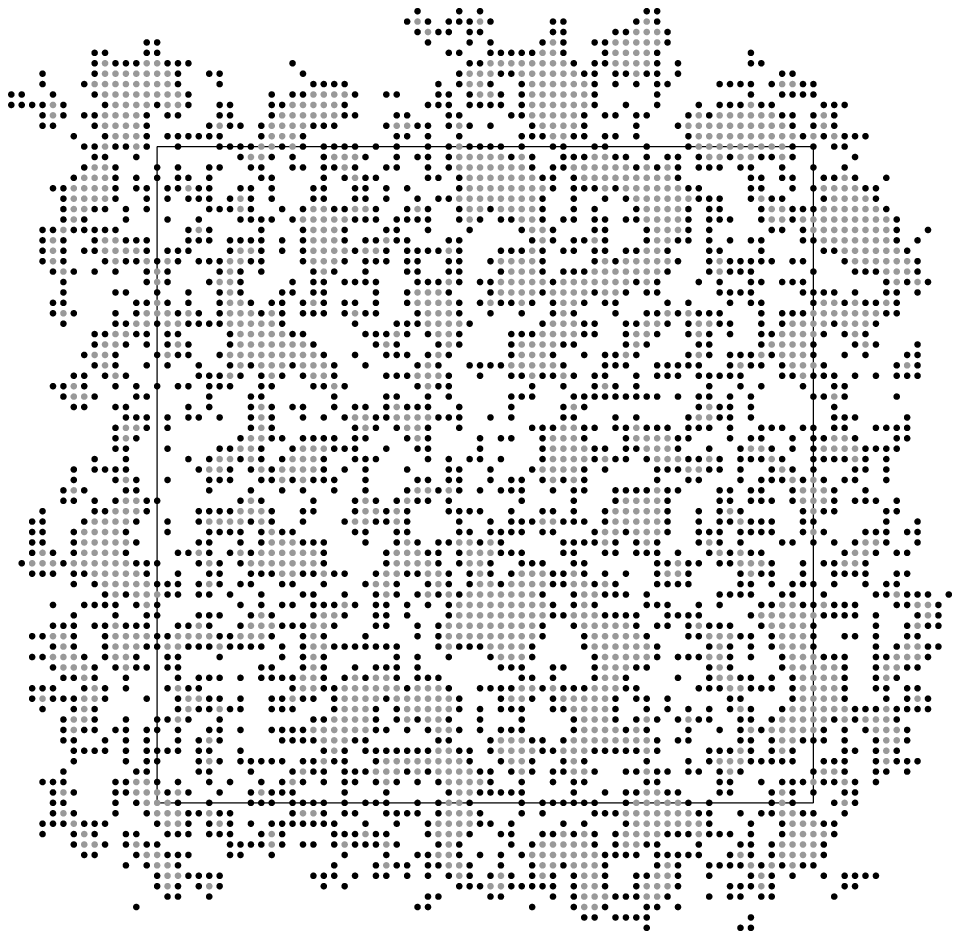,width=0.32\textwidth}
  \hfill\hfill\psfig{file=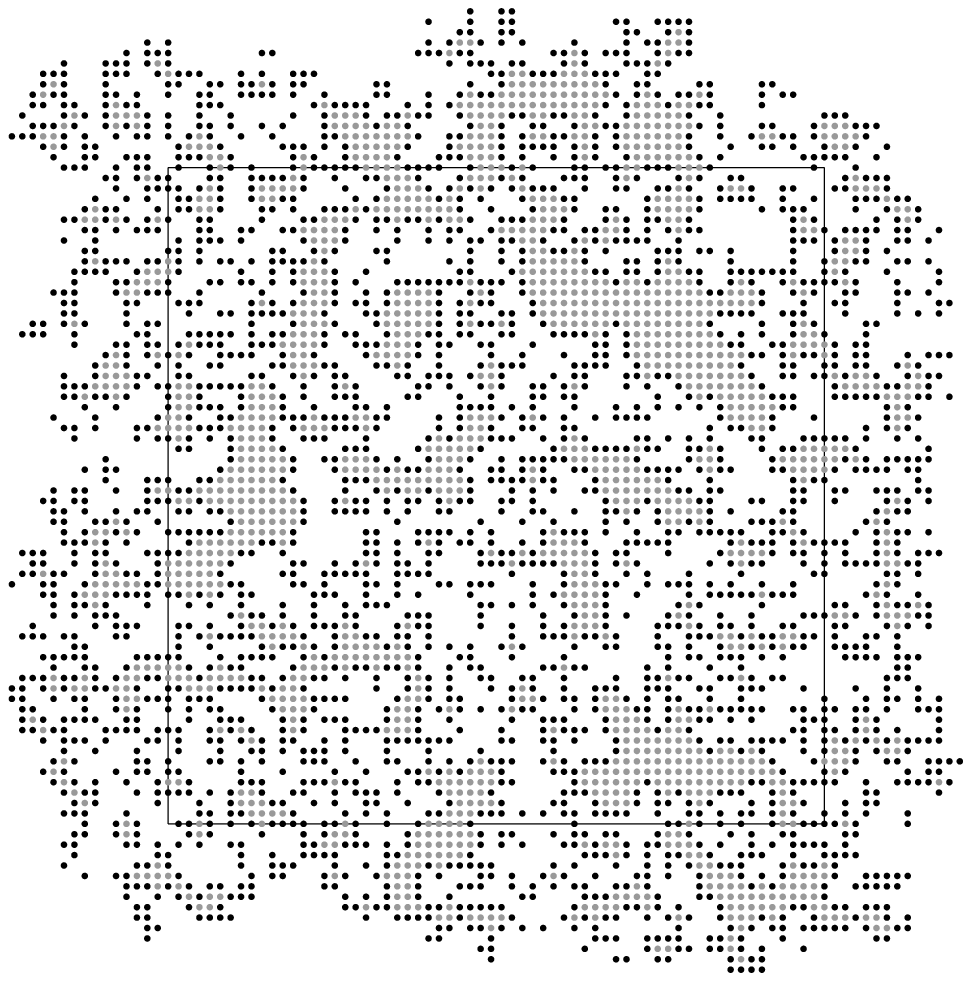,width=0.32\textwidth}
  \hfill}
\centerline{\hfill $(a)$ $t=2^{13}$ \hfill\hfill $(b)$ $t=2^{20}$
\hfill\hfill $(c)$ $t=2^{24}$ \hfill}
\caption{\label{pict} Morphology of the random solid system at three
  different times. Black dots are the externally accessible perimeter
  sites in contact with the solution. Gray dots are the interior solid
  sites, not in contact with the solution at that time. White space
  represents the solution. The solid line represents the lattice
  boundary at $t=0$.}
\end{figure}

Simulations have been performed on the square lattice of sizes ranging
from $L=8$ to $L=128$. Data are averaged over $\mathcal{N}_S=1000$ to
$10000$ samples. In figure \ref{pict}, time evolution of the system
morphology is shown for a system of size $L=64$ at three different
times $t=2^{12}$ $(a)$, $2^{20}$ $(b)$, and $2^{24}$ $(c)$.  It can be
seen that the system first becomes rough or porous, then the infinite
network of solid elements breaks into small finite clusters
(anti-percolation or dissolution) and finally equilibrates to a
morphology which remains almost unaltered over a long period of
time. A cluster of size $s$ contains $s$ number of atoms or molecules
connected by nearest neighbour bonds. The time evolution of the system
then has three different regimes, initial - porous, intermediate -
dissolution and final - dynamical equilibrium. There is also slow
dispersion of the clusters radially outward. It should be mentioned
here that in the study of self-stabilized etching of random solids by
finite etching solution (infinite solid and finite
solution)\cite{epl_physa} final stable morphology obtained was fractal
as it was observed by Bal\'azs in the corrosion of thin metal
films\cite{balazs}. In the following, different transitions will be
identified and the final equilibrium state of the system will be
characterized.

\begin{figure}
\centerline{\hfill \psfig{file=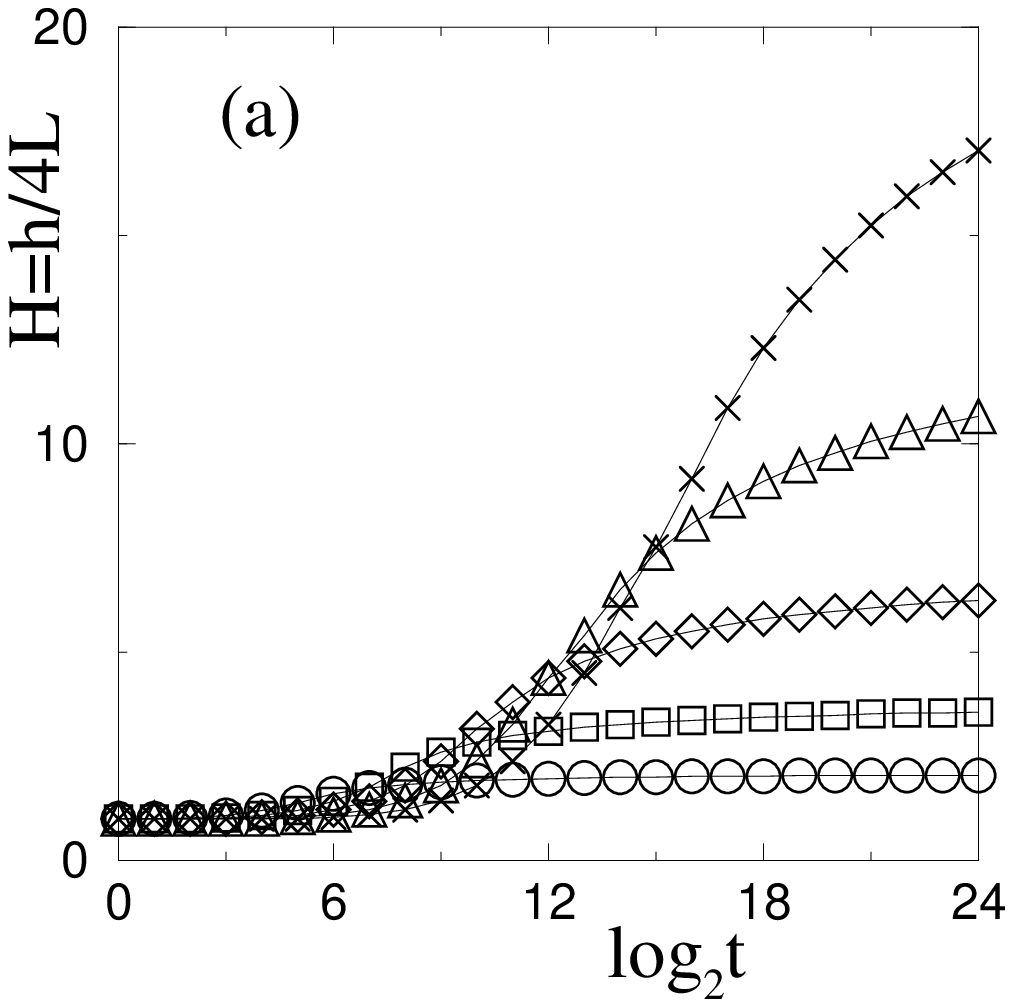,width=0.33\textwidth} \hfill
\psfig{file=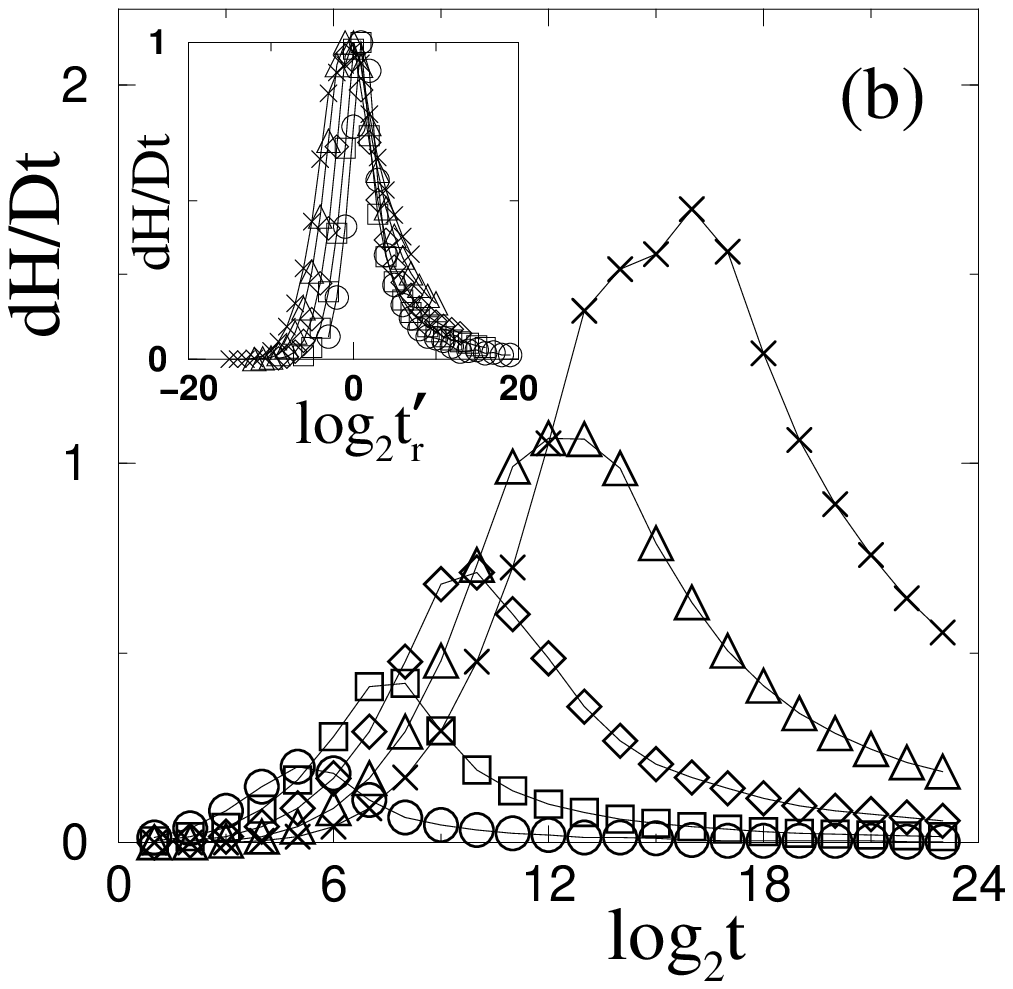,width=0.33\textwidth} \hfill
\psfig{file=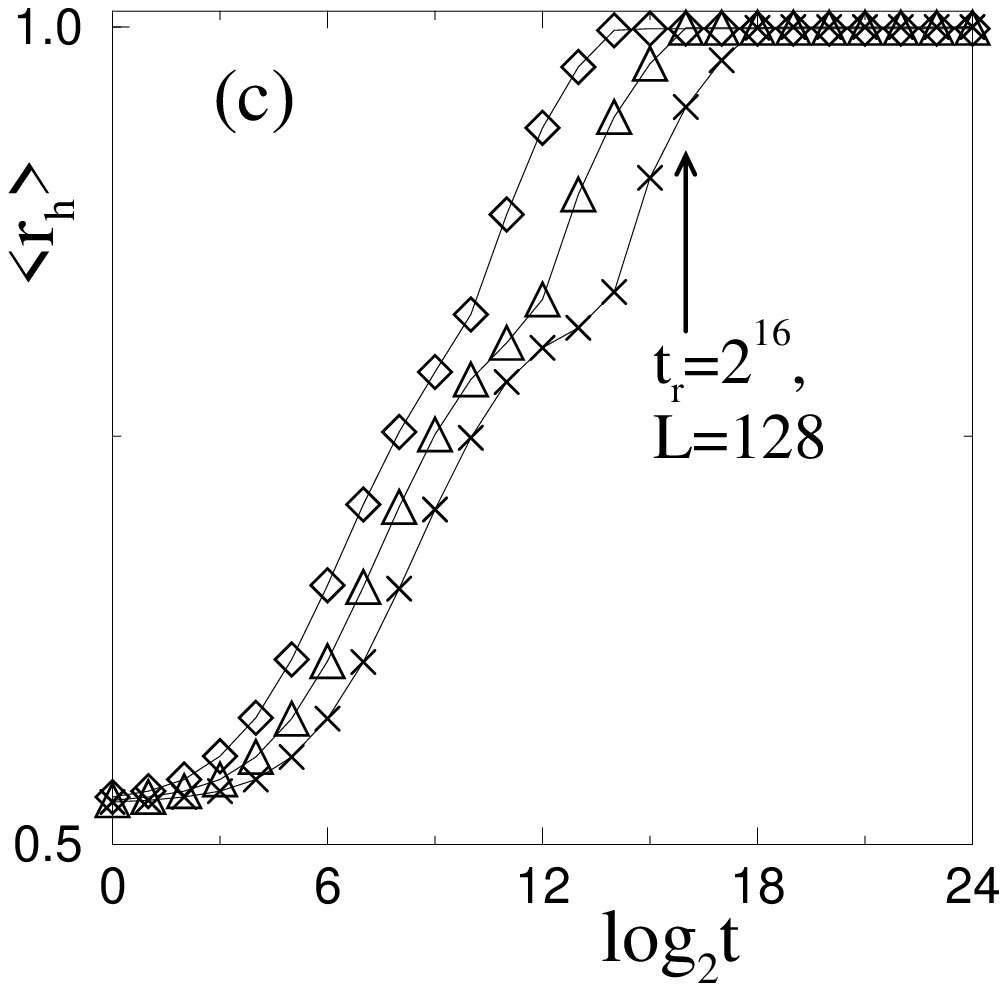,width=0.33\textwidth} \hfill }
\label{hull} 
\caption{$(a)$ Normalized hull size $H=h/4L$ and $(b)$ its logarithmic
time derivatives $dH/Dt$ ($Dt=d\log_2t$) are plotted against time
$t$. The symbols are: circles for $L=8$, squares for $L=16$,
diamonds for $L=32$, triangles for $L=64$ and crosses for $L=128$. In
the inset of ($b$), $dH/Dt$ is plotted against scaled roughening time
$t_r^\prime$. The height of the plots are adjusted to unity. Data for
all the systems collapse reasonably around $t_r^\prime =1$. $(c)$ Plot
of $\langle r_h\rangle$, average random numbers of the interface,
against $t$.}
\end{figure}

To determine the roughening or porosity transition, number of
externally accessible perimeter (hull) sites $h$ is counted with
time. The externally accessible perimeter sites, defined by Grossman
and Aharony\cite{gross}, include the sites available to the solution
molecules, touching the occupied sites of the clusters available at
that time. There are some standard methods for determining the
external perimeter or cluster hull like Ziff walk, kinetic
walk\cite{ziff} etc. However, instead of performing some walks around
the individual clusters, the time evolution of the hull itself could
be followed. At $t=0$, a string of lattice site index of all the
boundary sites is considered. The site with minimum random number is
taken out of the string and the string is locally modified depending
on the local environment in the original lattice. A string index is
then randomly chosen. A new random number is included at that position
and the string is locally modified again according to the local
environment in the original lattice. The hull string is then updated
at each time step. The total number of particles stored in this string
is the measure of the hull size. Note that, the string could be a
collection of hulls of all disconnected finite clusters and may not
represent a continuous path of perimeter sites of a single cluster. In
figure \ref{hull}($a$), normalized perimeter size $H=h/4L$ is plotted
against $t$. Initially $H$ increases slowly with time and finally
saturates followed by a rapid change in between. The roughening
transition corresponds to the maximum time rate of change of $H$. The
logarithmic time derivatives $dH/Dt$ ($Dt=d\log_2t$) are plotted in
figure \ref{hull}($b$) against $t$. It could be checked that the
maxima of $dH/Dt$ correspond to the highly porous morphology of the
system in figure \ref{pict} at $t=2^{13}$ for $L=64$. Notice that the
infinite network of solid elements still exists at this time. As the
roughening time $t_r$ is lattice size dependent, a scaling relation
for $t_r$ with $L$ is proposed as $t_r = 2^{-5}\times L^3$. Note that,
$t_r$ increse as $L^3$ and not as $L^2$, total number of
particles. The derivatives $dH/Dt$ are now plotted against the scaled
roughening time $t_r^\prime = t_r/(2^{-5}\times L^3)$ in the inset of
figure \ref{hull}($b$). Though there is not a good collapse of data
for $t_r^\prime < 1$ but it is clear that the transition occurs around
$t_r^\prime = 1$ independent of system size. It is important now to
consider the time evolution of the random numbers $r_h$ of the solid
elements on the hull. In figure \ref{hull}($c$), $\langle r_h\rangle =
\sum_i^hr_i/h$ is plotted against $t$. As the system evolves, $\langle
r_h\rangle$ increases and saturates to unity. However, the dynamics of
dissolution and redeposition will continue with the prefixed flux rate
since by definition a site with lowest random number will always
dissolve. The dynamics is not only independent of the absolute value
of $r_i$ but also independent of its distribution. For example, random
solid with random numbers uniformly distributed between $0.5$ to $1$
will not have different dynamics. It is interesting to notice that the
evolution of $\langle r_h\rangle$ has a plateau just before the
roughening transition and then it evolves again to another state of
constant $\langle r_h\rangle$. Before the roughening transition the
evolution then slows down temporarily and evolves again with higher
speed to another steady state. The plateau in $\langle r_h\rangle$
could represent a ``pseudo equilibrium'' corresponding to a transient
state. Appearance of pseudo equilibrium was also observed in the
dissolution study of multicomponent random solid\cite{santra}.

The present model of interface evolution has great similarity with the
Bak-Snappen (BS) model of biological evolution\cite{bs}. In the BS
model, time evolution of the fitness string of a number of species was
considered. The fitness of a species was represented by a random
number uniformly distributed between $0$ and $1$. The species with
lowest fitness (lowest random number) was replaced by another random
number and the fitness string was locally modified as the hull is
modified locally here in the present problem. The BS model
self-organizes into a critical state with intermittent co-evolutionary
avalanches of all sizes representing punctuated equilibrium behaviour
of the evolution process. The average random number of the hull
equivalent to the global fitness shows only one plateau during the
time evolution of the interface. The fitness string in the BS model
never had a hole, always a single continuous string, unlike the hull
here which could be a collection of external perimeters of all finite
disconnected clusters.

\begin{figure}
\centerline{\hfill \psfig{file=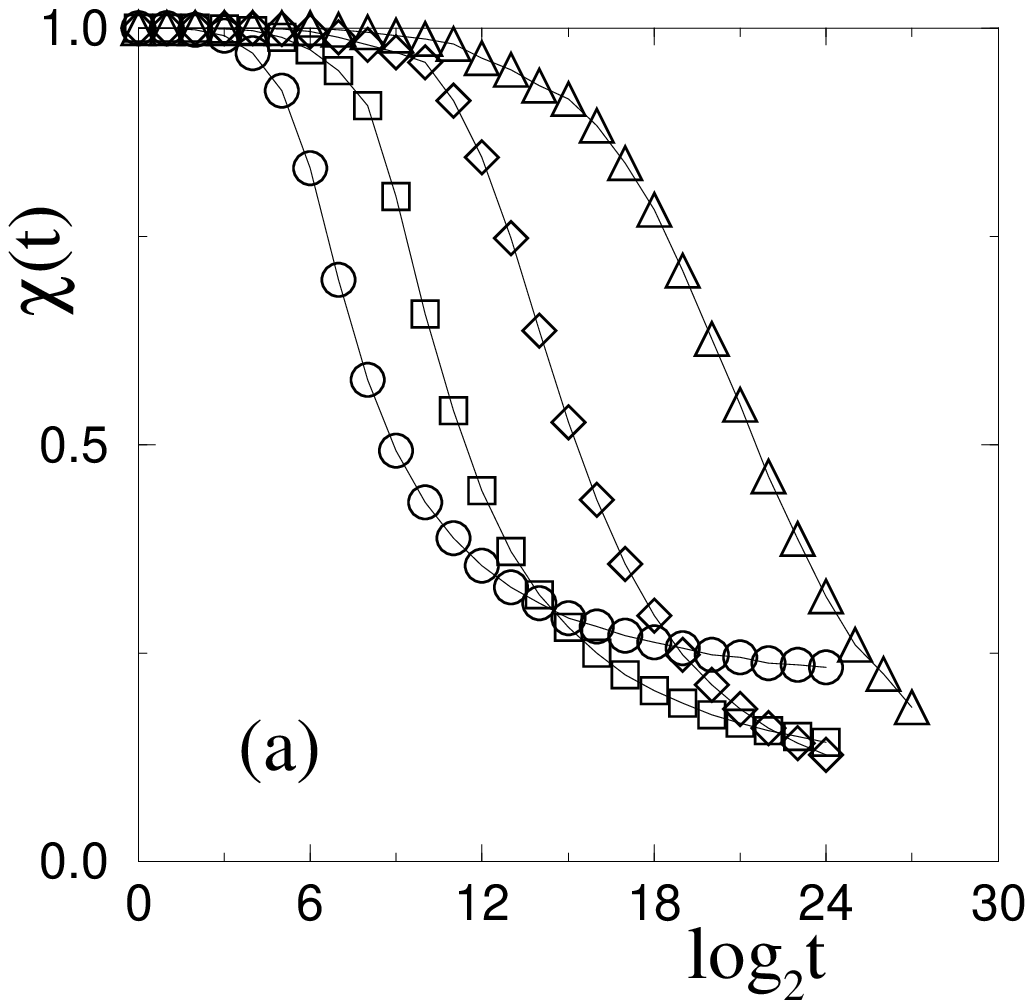,width=0.33\textwidth}
\hfill \psfig{file=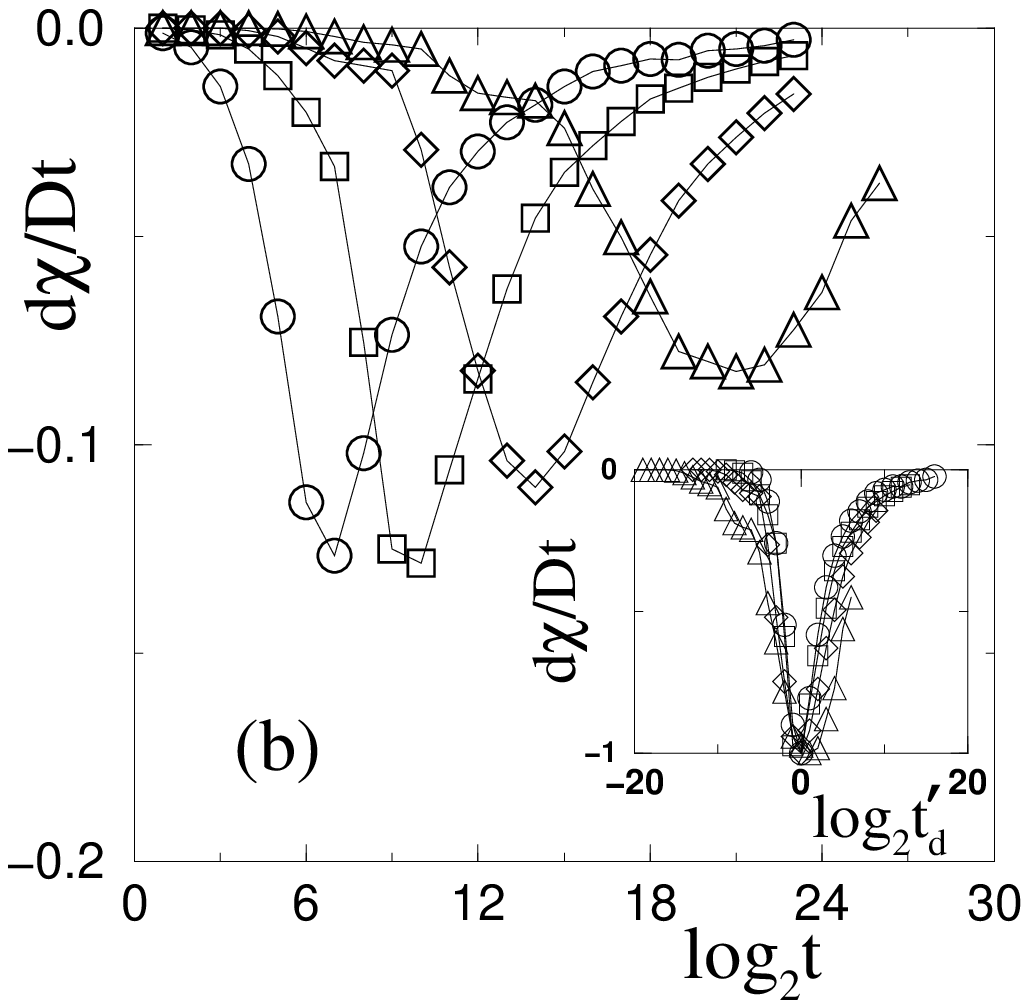,width=0.33\textwidth}\hfill
\psfig{file=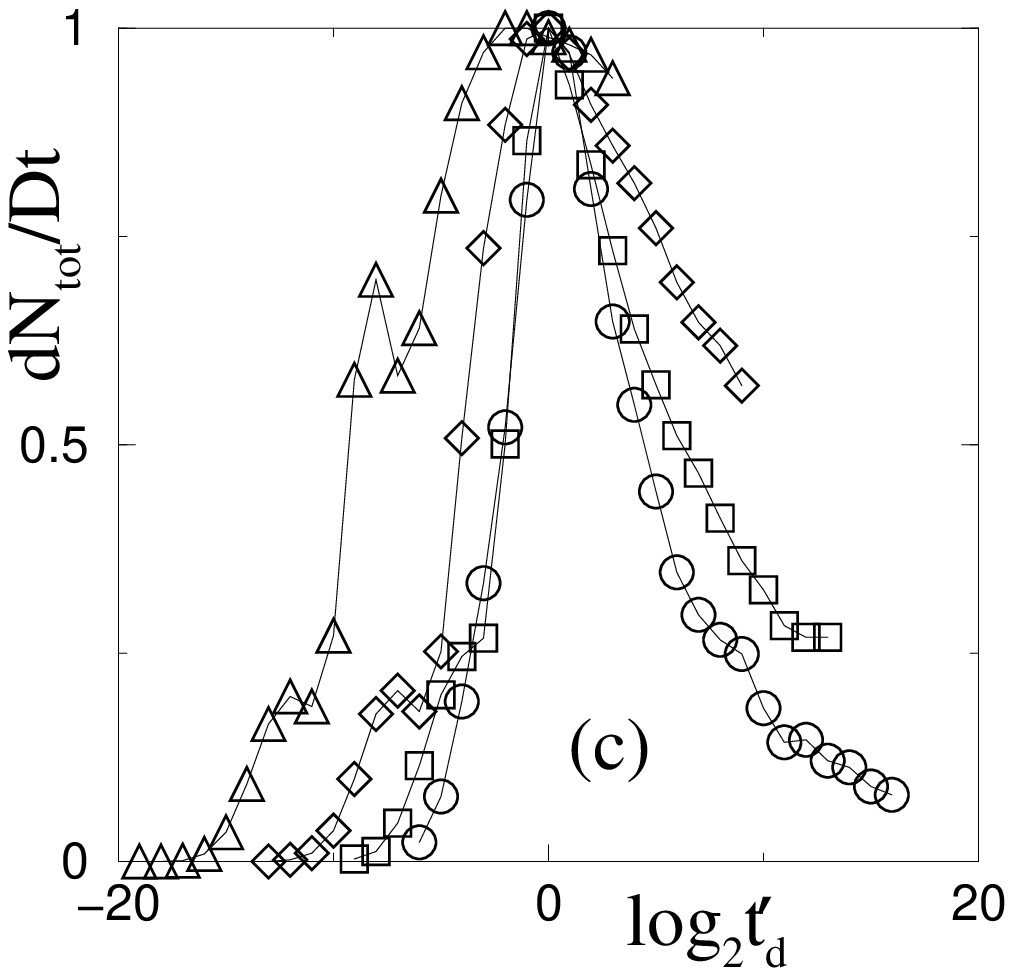,width=0.33\textwidth}\hfill }
\caption{Plot of (a) average cluster size $\chi(t)$, (b) its
  logarithmic time derivative $d\chi/Dt$ ($Dt=d\log_2t$) against
  $t$. $\chi$ is normalized to unity dividing by the total number of
  particles $L^2$. In the inset of ($b$), $d\chi/Dt$ is plotted
  against scaled dissolution time $t'_d$. The height of the plots are
  normalized to unity. ($c$) Plot of $dN_{tot}/Dt$ against
  $t_d^\prime$ indicates maximum cluster generation at
  $t_d^\prime=1$. The same symbol set for different $L$ of the
  previous figure is used. For $L=128$, dissolution does not occur by
  $t=2^{24}$ time steps.}
\label{clssz} 
\end{figure}

Far from the roughening transition, the dissolution or
anti-percolation transition occurs in the system when $\langle
r_h\rangle$ saturates to unity. At the anti-percolation transition,
the infinite network of the random solid breaks into small finite
clusters for the first time. The probability to have an $s$-sited
cluster at time $t$ is given by
\begin{equation}
\label{probs}
P_s(t) = n_s(t)/N_{tot}(t)
\end{equation}
where $n_s(t)$ is the number of $s$-sited cluster out of total
$N_{tot}$ clusters at that time. To identify the anti-percolation
transition, average cluster size $\chi(t)=\sum s^2P_s(t)/\sum sP_s(t)$
is calculated as function of time. Note that the sum is over all
possible clusters including the largest cluster, unlike
percolation\cite{perco}. $\chi(t)$ is plotted against $t$ in figure
\ref{clssz}($a$). It can be seen that the average cluster size remains
almost constant initially then decreases rapidly during the
anti-percolation transition and finally saturates to a small
value. Dissolution time $t_d$ is determined from the maximum time rate
of change of $\chi(t)$. The logarithmic time derivative $d\chi/Dt$ of
the average cluster size is plotted against $t$ in figure
\ref{clssz}($b$). The dips in the plots correspond to the
anti-percolation or dissolution at which the infinite network
disappears from the system for the first time. The corresponding
morphology is shown in figure \ref{pict}($b$) at $t=2^{20}$ for
$L=64$. A scaling form $t_d = L^2\times 2^{L/8}$ is suggested for $L$
dependent dissolution time.  $d\chi/Dt$ is now plotted against the
scaled dissolution time $t_d^\prime= t_d/(L^2\times 2^{L/8})$ in the
inset of figure \ref{clssz}($b$). A reasonable data collapse is
observed. The system then undergoes an anti-percolation transition at
$t_d^\prime=1$ independent of system size. The dissolution time $t_d$
can also be estimated from the total number of clusters $N_{tot}(t)$
generated with time. The time rate of change of $N_{tot}$,
$dN_{tot}/Dt$, is plotted in figure \ref{clssz}($c$) for different
lattice sizes. The maximum cluster generation occurs at $t=t_d$ and
then the rate of cluster generation decreases. Notice that roughening
time $t_r$ and dissolution time $t_d$ scales very differently, $t_r$
increases as $L^3$ whereas $t_d$ increases exponentially with $L$.

\begin{figure}
\centerline{\hfill \psfig{file=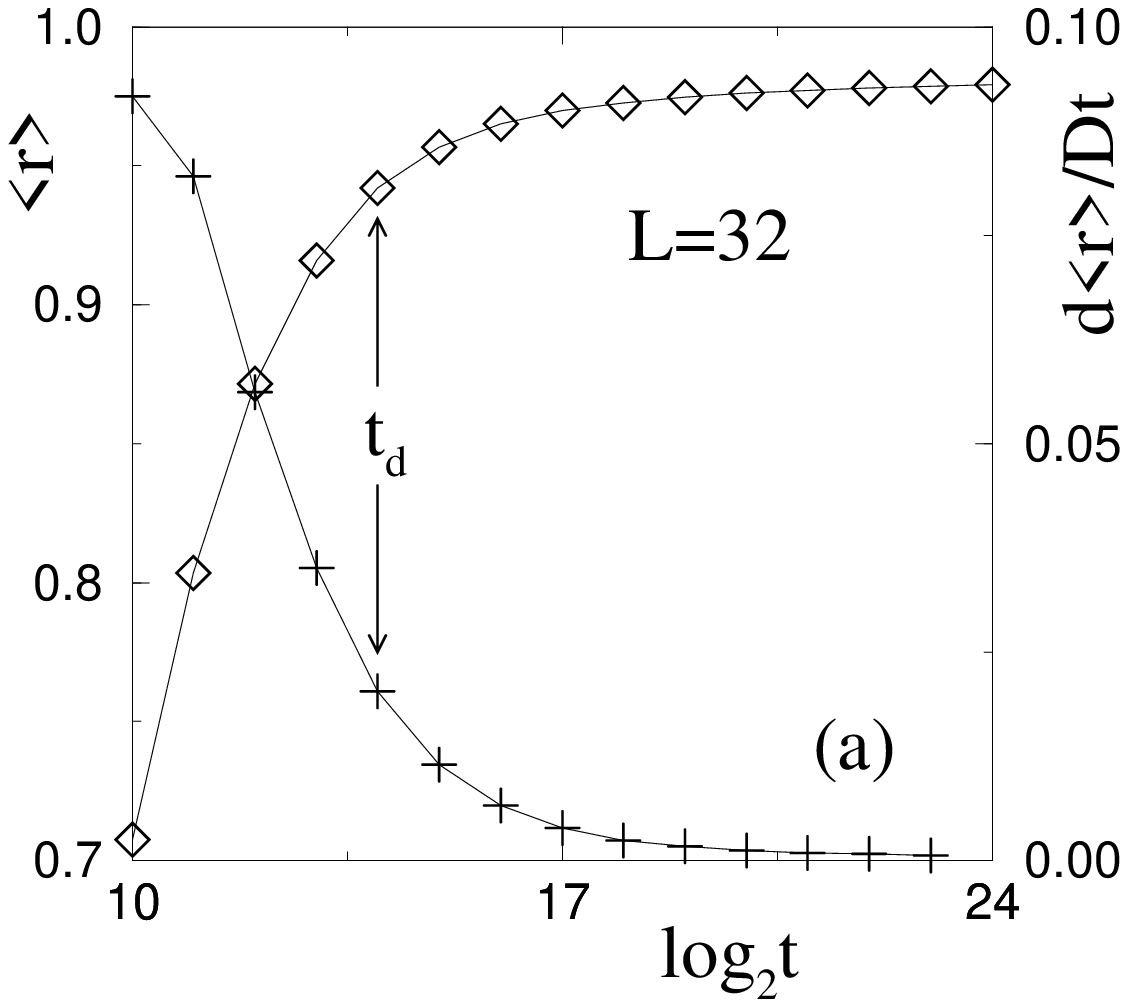,width=0.38\textwidth}
\hfill \psfig{file=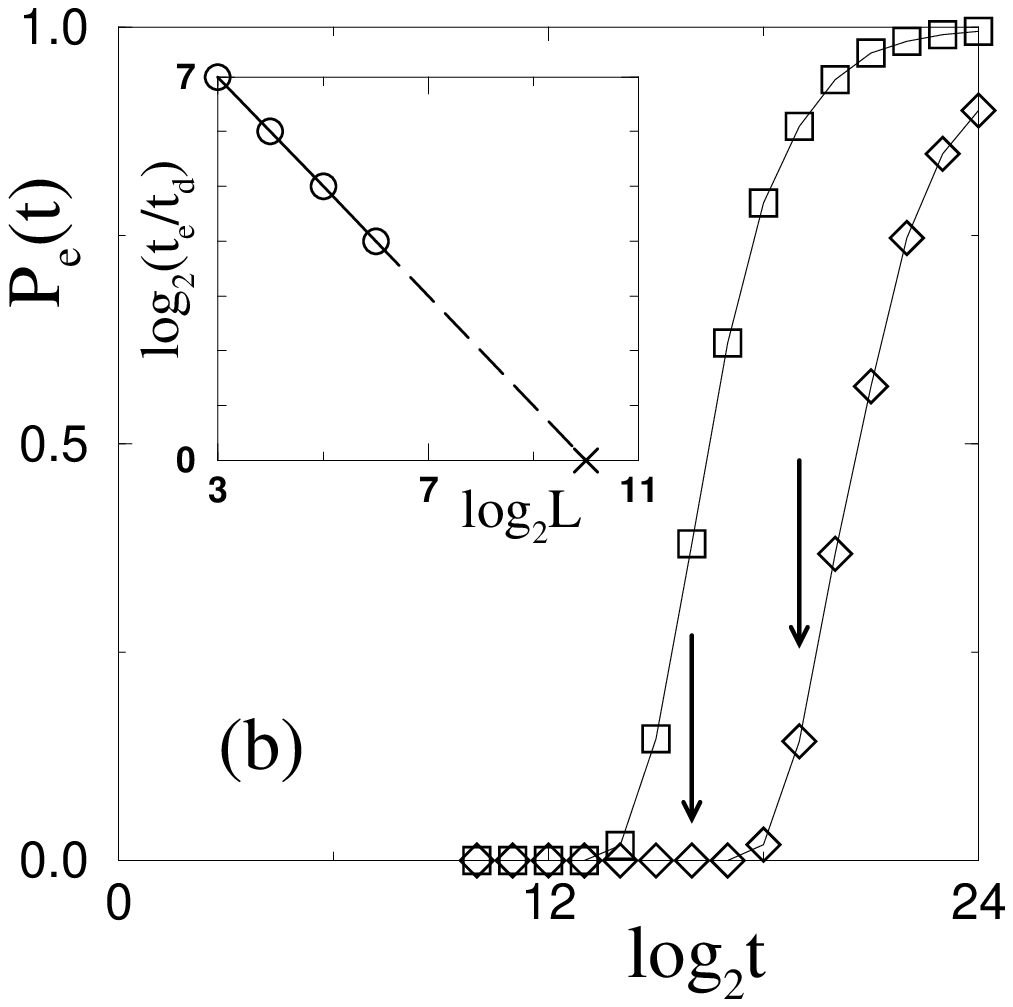,width=0.325\textwidth} \hfill
\psfig{file=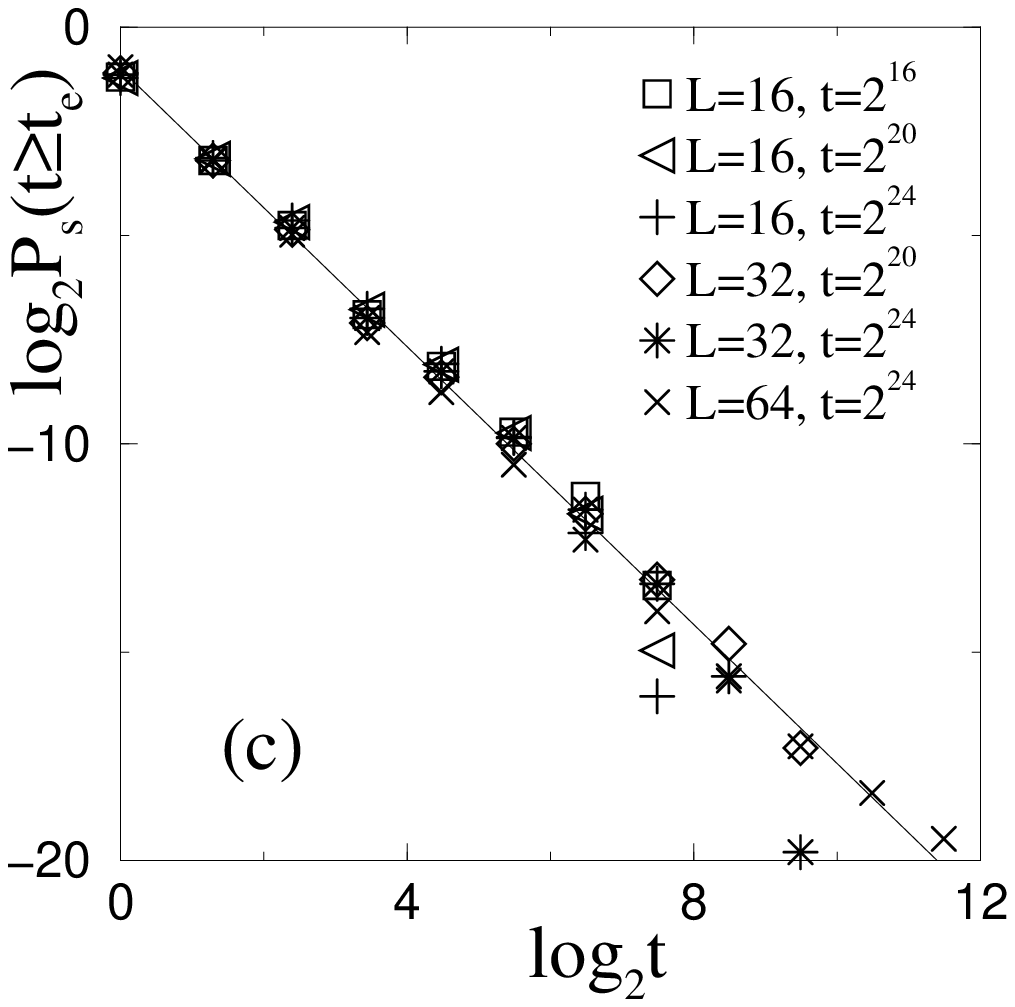,width=0.33\textwidth} \hfill}
\caption{\label{csdti} ($a$) Plot of $\langle r\rangle$ ($\diamondsuit$)
  and $d\langle r\rangle/Dt$ ($+$) against $t$ for $L=32$. Arrows
  indicate dissolution time. ($b$) Plot of $P_e(t)$ for $L=16$
  ($\Box$) and $32$ ($\diamondsuit$). In the inset, scaling of $t_e/t_d$
  with $L$ is shown. Difference between $t_d$ and $t_e$ vanishes at
  $L=2^{10}$. ($c$) Plot of cluster size distributions $P_s(t)$ at
  different times for $t\ge t_e$ for different lattice sizes. Symbols
  are explained in the legend. The straight line with slope $\approx
  1.67$ is guide to eye.  }
\end{figure}

The system is then allowed to evolve further, far from the
anti-percolation transition. As $t\rightarrow \infty$, it is already
seen that hull size $h$ and average cluster size $\chi$ are going to
saturate to some finite values. Now it is important to consider the
evolution of all the random numbers instead that of only the
hull. Average random number $\langle r\rangle = \sum_i^{L^2}r_i/L^2$
of all the sites and $d\langle r\rangle/Dt$ are plotted against $t$ in
figure \ref{csdti}($a$) for $L=32$ starting from porosity transition
$t_r=2^{10}$. Notice that $\langle r\rangle$ saturates to one just
after the anti-percolation transition and $d\langle r\rangle/Dt$ goes
to zero. Since $\langle r\rangle$ is close to one, it is expected that
most of the particles have changed their random numbers or dissolved
at least once. The probability $P_e(t)$ to have a sample at time $t$
with all the sites dissolved at least once is calculated. It is
defined as $P_e(t) = \mathcal{N}_d(t)/\mathcal{N}_S$ where
$\mathcal{N}_d(t)$ is the number of samples for which all the sites
dissolved at least once by time $t$ out of $\mathcal{N}_S$ samples. It
is plotted in figure \ref{csdti}($b$) against $t$ for $L=16$ and
$32$. A time $t_e$ is defined corresponding to the maximum change in
$P_e$ at which all the sites of $50\%$ samples dissolved at least
once. It is obtained as $t_e=2^{14}$ for $L=8$, $t_e=2^{16}$ for
$L=16$, $t_e=2^{19}$ for $L=32$ and $t_e=2^{24}$ for $L=64$. It is
interesting to notice that the logaritghmic difference in $t_e$ and
$t_d$ is decreasing with the system size. $\log_2(t_e/t_d)$ is plotted
in the inset of ($b$) with $\log_2L$. The difference vanishes at
$L=2^{10}$. Thus, beyond the system size of $2^{10}$ all the sites
will dissolve at least once by the dissolution time $t_d$. The
situation can be called as complete dissolution. At this stage, the
small finite clusters in the solution can be considered as a single
fluidized particle phase of the system and the dynamics is
fragmentation of a particle in one cluster and coagulation at another
cluster with a slow radially outward dispersion of the
clusters. Constant average cluster size $\chi$ suggests that the
cluster size distribution remains invariant over time at this
phase. It is verified by plotting the cluster size distribution
$P_s(t)$ at $t\ge t_e$ for different $L$ in figure
\ref{csdti}($c$). The distributions collapse onto a single curve for
different $t$ and $L$. Note that for $L=16$, the distribution remains
invariant for $t=2^{16}(t_e)$ to $t=2^{24}$. The distribution follows
a power law with an exponent $\tau^\prime \approx 1.67$. The same
slope is also obtained for $P_s=n_s/L^2$ distribution, as in
percolation, since after dissolution time $t_d$ there will be very few
cluster generation for large systems. It is then a critical state of
the system with no characteristic size. After anti-percolation, the
system then sattels into a state of constant cluster size distribution
through critical slowing down ($d\langle r\rangle/Dt=0$). The final
fluidized particle phase of the system then can be considered as a
dynamical equilibrium state characterized by constant chemical
potential (saturated hull size) and constant average cluster size. The
transitions observed here during time evolution are features of slow
dynamics considered in the system. Inclusion of rapid dynamics in the
system could musk all these transitions and the system may evolve
directly to the final equilibrium state.

The anti-percolation transition discussed above is very similar to
fragmentation phenomena in $2D$. In the case of fragmentation, the
system equilibrates after complete dissipation of the impact force and
a power law distribution of the fragment mass ($m$) was obtained
$i.e.$; $n(m)\sim m^{-\tau}$. The critical behaviour of fragmentation
was studied by Kun and Herrmann\cite{hans} and found that it belongs
to the same universality class of percolation, $i.e.$;
$\tau=187/91$\cite{perco}. However, in a recent experiment of
fragmentation of brittle solids by Katsuragi et al\cite{katsu} it is
found that $\tau\approx 1.5$, different from percolation. The exponent
obtained here ($1.67$) for the cluster size distribution at the
anti-percolation transition is different from both Katsuragi exponent
in the case of fragmentation as well as that of percolation.

In conclusion, the chemical reaction of dissolution of a random solid
in a solution is modeled. The model demonstrates different transitions
like roughening or porosity, dissolution or anti-percolation during
time evolution. The transitions physically correspond to the maximum
solid-liquid interface and disappearance of the infinite cluster
respectively. Different transition times are found scaling very
differently with the system size. Finally the system evolves to a
dynamical equilibrium state through critical slowing down charcterized
by constant cluster size distribution. The dynamical system considered
here, driven by the initial difference in chemical potential,
self-organizes itself into a critical state with no characteristic
size at dynamical equilibrium. This is a new phenomenon. Generally,
the dynamical systems, driven externally, self-organizes themselves to
a non-equilibrium steady state characterized by power law correlations
as observed in SOC models. The phenomena observed here should exists
in $3D$ systems. The present model has relevance with other widely
different models like biological evolution or fragmentation of brittle
solids. The model could also be useful in understanding time evolution
of waste disposal glasses.

The authors thank B. Sapoval, I. Bose and A. Srinivasan for critical
comments. SS thanks CSIR, India for financial support.

\end{document}